\documentclass[aip,apl,twocolumn,reprint,preprintnumbers,amsmath,amssymb]{revtex4-1}

\usepackage{color}
\usepackage{graphicx}
\usepackage{dcolumn}
\usepackage{bm}

\newcommand{\sqrtwo}{\ensuremath{\sqrt{2}}}

\begin{document}

\title{Tuning the Curie temperature of FeCo compounds by tetragonal distortion}

\author{A. Jakobsson}
\affiliation{Peter Gr\"{u}nberg Institut and Institute for
Advanced Simulation, Forschungszentrum J\"{u}lich and JARA, 52425
J\"{u}lich, Germany} \affiliation{Department of Physics and
Astronomy, Uppsala University, Box 516, 75120 Uppsala, Sweden}

\author{E.
\c{S}a\c{s}{\i}o\u{g}lu}\email{e.sasioglu@fz-juelich.de}
\affiliation{Peter Gr\"{u}nberg Institut and Institute for
Advanced Simulation, Forschungszentrum J\"{u}lich and JARA, 52425
J\"{u}lich, Germany}

\author{Ph. Mavropoulos}
\affiliation{Peter Gr\"{u}nberg Institut and Institute for
Advanced Simulation, Forschungszentrum J\"{u}lich and JARA, 52425
J\"{u}lich, Germany}

\author{M. Le\v{z}ai\'{c}}
\affiliation{Peter Gr\"{u}nberg Institut and Institute for
Advanced Simulation, Forschungszentrum J\"{u}lich and JARA, 52425
J\"{u}lich, Germany}

\author{B. Sanyal}
\affiliation{Department of Physics and
Astronomy, Uppsala University, Box 516, 75120 Uppsala, Sweden}

\author{G. Bihlmayer}
\affiliation{Peter Gr\"{u}nberg Institut and Institute for
Advanced Simulation, Forschungszentrum J\"{u}lich and JARA, 52425
J\"{u}lich, Germany}

\author{S. Bl\"{u}gel}
\affiliation{Peter Gr\"{u}nberg Institut and Institute for
Advanced Simulation, Forschungszentrum J\"{u}lich and JARA, 52425
J\"{u}lich, Germany}

\begin{abstract}

Combining density-functional theory calculations with a classical
Monte Carlo method, we show that for B2-type FeCo compounds
tetragonal distortion gives rise to a strong reduction of the
Curie temperature $T_{\mathrm{C}}$. The $T_{\mathrm{C}}$
monotonically decreases from 1575 K (for $c/a=1$) to 940 K (for
$c/a=\sqrtwo$). We find that the nearest neighbor Fe-Co exchange
interaction is sufficient to explain the $c/a$ behavior of the
$T_{\mathrm{C}}$. Combination of high magnetocrystalline
anisotropy energy with a moderate $T_{\mathrm{C}}$ value suggests
tetragonal FeCo grown on the Rh substrate with $c/a=1.24$ to be a
promising material for heat-assisted magnetic recording
applications.

\end{abstract}


\maketitle

The recording density in a commercial hard disk drive (HDD), i.e.,
the amount of information that can be stored per square inch, has
increased by more than 7 orders of magnitude since its first
introduction in 1956.\cite{IBM} Such an increase has been achieved
by a simple scaling of the dimensions of the bits recorded in
storage medium. However, the recording density has an upper limit
due to the superparamagnetic effect and limited write field of the
writing head. This limit is around 1 Tbit per square inch for
current perpendicular magnetic
recording.\cite{Fontana,Piramanayagam,Plos,Plumer} In order to
further increase the recording density in future recording media
new materials with the following properties are sought: (i) they
should have large uniaxial magneto-crystalline anisotropy energy
(MAE) $K_\mathrm{u}$, (ii) large saturation magnetization, (iii)
fast magnetic response to external applied fields, and (iv)
moderate Curie temperatures above the room temperature.

Retaining the magnetization of the medium over a long period of
time despite thermal fluctuations is one of the major problems in
designing magnetic storage media. If the ratio of the  magnetic
energy per grain $K_\mathrm{u}V$, where $V$ is the grain volume, to
the thermal energy $k_\mathrm{B}T$ becomes sufficiently small, the
thermal fluctuations can reverse the magnetization in a region of
the medium destroying the information stored there.\cite{Piramanayagam,Weller}
To further increase the recording density high-$K_\mathrm{u}$ materials  
with large saturation magnetization $M_\mathrm{s}$ are needed. The 
large $M_\mathrm{s}$ is beneficial to reduce the write field. In
addition to large $K_\mathrm{u}$ and $M_\mathrm{s}$ values,
another important issue in magnetic recording applications is the
magnetic switching time, which imposes physical limits on areal
recording densities and data rates.\cite{Plumer} In current
devices the switching speeds have reached a point where dynamical
effects are becoming important.\cite{Visscher,Safanov,Fidler,Garanin,Silva,Lenz,Baberschke}
Collective magnon excitations play an important role in fast
precessional magnetic switching processes because they serve as a 
heat bath for dissipation of the Zeeman energy and thus contribute to
the relaxation of magnetization and switching time.

On the other hand, the large-$K_\mathrm{u}$ materials require very
high magnetic fields for writing the information to the recording
media. As the bit size gets smaller and smaller, at some point the
magnetic field required for switching the magnetization direction
exceeds the maximal available magnetic writing fields and thus data
can no longer be written to the disk. To solve this problem the
heat-assisted magnetic recording (HAMR) was proposed as a promising
approach, which enables large increases in the storage density of 
HDD.\cite{Plumer,HAMR,HAMR2} In HAMR a laser is used to momentarily and
locally heat the recording area of the medium to reduce its
coercivity. It has been suggested that magnetic recording close to or
above the Curie temperature is required to achieve the highest areal
density advantage of HAMR, making the $T_{\mathrm{C}}$ an important
parameter for applications and choice of materials.\cite{Tc_1,Tc_2}
With increasing temperature the $K_\mathrm{u}$ of
the medium decreases and above $T_{\mathrm{C}}$ 
$K_\mathrm{u}$ vanishes, making it possible to write the information
with available head fields.  Thus, the $T_{\mathrm{C}}$ is an
important parameter in the design of HAMR media.

Materials that combine most of necessary conditions for HAMR
applications are B2-type tetragonal FeCo compounds. The large values
of $K_\mathrm{u}$, reaching 600~$\mu$eV, and $M_\mathrm{s}$ in these
compounds were first predicted by first-principles calculations\cite{Burkert} 
and then confirmed by experiments.\cite{Andersson,Luo,yildiz_1,yildiz_2}  Experimentally, FeCo
compounds have been grown on the Pd, Ir, and Rh substrates in B2-type
structure, in which the in-plane lattice constant $a$ is enforced by
the substrate and the out-of-plane lattice constant $c$ changes so as
to keep the volume constant.  In particular, Yildiz \textit{et
al.}\cite{yildiz_1,yildiz_2} found in agreement with theoretical
predictions that the perpendicular magnetic anisotropy is very
sensitive to the tetragonal distortion and increases with increasing
$c/a$ ratio, which allows to tune the perpendicular anisotropy value
by growing the alloys on different substrates.  Yildiz \textit{et al.}
\cite{yildiz_1,yildiz_2} have also shown that the structure remains
stable for film thicknesses of up to 15 monolayers.  Note also that
microscopic atomic order in B2-type FeCo compounds is crucial to
achieve high $K_\mathrm{u}$ values.\cite{Turek_CPA,Neise11}

At low temperatures the ordered cubic FeCo takes the CsCl (B2)
structure, at around 1000 K it undergoes an order-disorder transition
and at around 1230 K a bcc-fcc transformation accompanied by a
magnetic-nonmagnetic transition.\cite{FeCo_CsCl_1,FeCo_CsCl_2,Bozorth}
In a recent paper \c{S}a\c{s}{\i}o\u{g}lu \emph{et al.}\cite{Sasioglu_2013} have
studied the effect of the tetragonal distortion on magnon spectra of
the B2-type FeCo compounds by employing the many-body perturbation
theory. The authors have shown that tetragonal distortion gives rise
to a significant magnon softening, which indicates a strong reduction
of the Curie temperature. Ab-initio calculations by Le\v{z}ai\'{c}
\emph{et al.}\cite{Lezaic} on cubic ordered and disordered FeCo
alloys have shown that the calculated $T_{\mathrm{C}}$ agrees well
with experiment.

\begin{figure*}[!ht]
\includegraphics[scale=0.73]{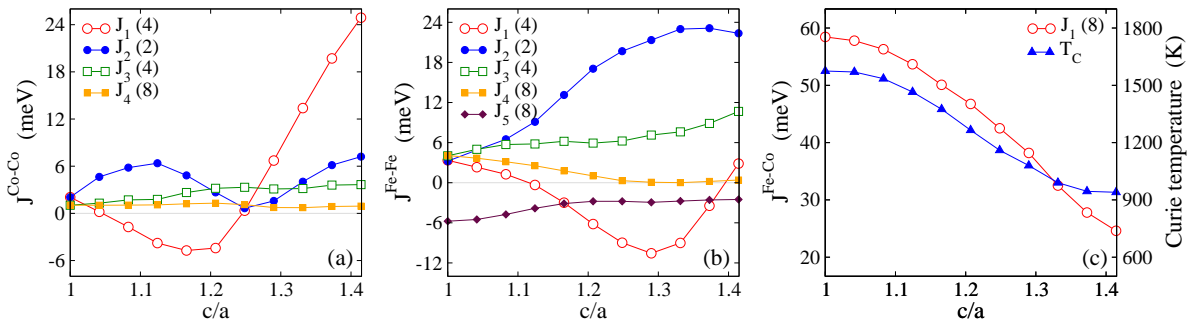}
\vskip -0.2 cm 
\caption{(Color online) (a) First four nearest neighbor
intra-sublattice Co-Co exchange parameters as a function of
tetragonal distortion in B2-type FeCo. (b) the same for Fe
sublattice for the first five shells. (c) Nearest neighbor
inter-sublattice Fe-Co exchange parameters and estimated Curie
temperature of the FeCo compounds as a function of tetragonal
distortion. In each panel for each exchange parameter the number of
atoms in the corresponding coordination sphere is given. Positive
exchange parameters correspond to ferromagnetic coupling, negative
to antiferromagnetic.}\label{fig1}
\end{figure*}

The aim of this Letter is to study the effect of the tetragonal
distortion on the Curie temperature of the B2-type FeCo compounds from
first principles. It is shown that tetragonal distortion gives rise to
a strong reduction of $T_{\mathrm{C}}$, that decreases from 1575 K
(for $c/a=1$) to 940 K (for $c/a=\sqrtwo$).  Combination of moderate
$T_{\mathrm{C}}$ values together with large $K_\mathrm{u}$ suggests
B2-type tetragonal FeCo grown on the Rh substrate with $c/a=1.24$ to
be a promising material for HAMR applications.

We calculate the Curie temperature using an established approach: the
adiabatic approximation for the calculation of magnon
spectra.\cite{adiabatic_1,adiabatic_2,Adam_2013} {\em Ab initio}
total-energy results, calculated within the frozen-magnon
approximation, are mapped to the classical Heisenberg model,
\begin{equation}
H=-\frac{1}{2}\sum_{i,j(i\ne
j)}J_{ij}{\mathbf{e}_{i}}\cdot{\mathbf{e}_{j}}\,,\label{eq:1}
\end{equation}
where $J_{ij}$ are the exchange constants between the magnetic moments
at sites $i$ and $j$ and ${\mathbf{e}_{i}}$ is a unit vector along the
moment of atom $i$. Accounting for the interactions up to the 12th
nearest neighbor, $T_{\mathrm{C}}$ is calculated within this model
by a Monte Carlo method by locating the crossing point of the
fourth-order cumulants\cite{Landau00} for 5488 and 8192-atom
supercells in the bulk limit and neglecting the anisotropy, which is a
good approximation for a film thickness of 15 monolayers.\cite{Bulk,Erickson91}

The ab initio results are calculated within the generalized gradient
approximation\cite{GGA} to density-functional theory. We employ the
full-potential linearized augmented plane-wave (FLAPW) method as
implemented in the \texttt{FLEUR} code.\cite{Fleur} The muffin-tin
radii of Fe and Co are chosen to be 1.21 \AA. A dense
$16\times16\times16$ $\mathbf{k}$-point grid is used.  Keeping the
volume of the B2-type unit cell constant ($V=23.766$ \AA$^3$) we vary
the $c/a$ ratio from 1 to \sqrtwo.  Note that if both atoms in the
unit cell were identical we would get a bcc lattice for $c/a=1$ and
fcc for $c/a=\sqrtwo$.  As FeCo with $c/a=1$ crystallizes in ordered
B2 structure\cite{FeCo_CsCl_1,FeCo_CsCl_2} we assume the same type of
structure with additional tetragonal distortion in our calculations.
The mechanism behind the giant uniaxial MAE observed in tetragonal
FeCo compounds has been discussed in detail in
Ref.\,\onlinecite{Burkert} and will not be analyzed here. Indeed, our
calculated values of uniaxial MAE (results not shown) are very similar
to those reported by Burkert \textit{et al.}\cite{Burkert}

\begin{figure}
\includegraphics[width=0.85\columnwidth]{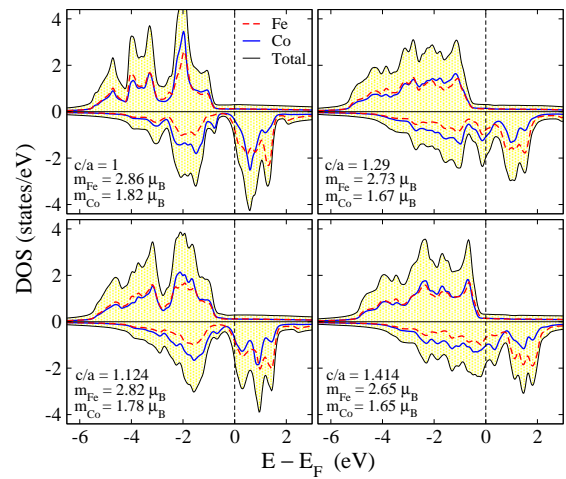}
\vskip -0.2 cm \caption{(Color online) Total and atom-resolved
density of states of B2-type FeCo compounds for four different
$c/a$ ratios. In each panel we include atom-resolved magnetic
moments. The positive (negative) DOS axis corresponds to the majority-spin
(minority-spin) channel.} \label{fig2}
\end{figure}

We begin the discussion of our results by presenting the exchange
interactions.  Figures \ref{fig1}(a) and (b) show the calculated
sizeable intra-sublattice Co-Co and Fe-Fe exchange parameters,
respectively, as a function of tetragonal distortion, i.e., from
$c/a=1$ to $c/a=\sqrtwo$ (more distant coupling parameters are
included in the $T_{\rm C}$ calculation but not shown
here). Figure\,\ref{fig1}(c) shows the $c/a$ dependence of the nearest
neighbor inter-sublattice Fe-Co exchange parameters as well as the
calculated Curie temperature of the compounds. Due to the strong
ferromagnetic nature of FeCo compounds (very low majority-spin DOS,
see Fig.\,\ref{fig2}) the absolute value of the exchange parameters
decays quickly with increasing interatomic distance\cite{Pajda} and
the main contribution to $T_{\mathrm{C}}$ comes from the interaction
between atoms lying in a distance of a few first neighboring
shells. The importance of each interaction ($J_1$, $J_2$, etc.) should
be judged taking into account the number of neighbours in the
corresponding coordination sphere, given in parentheses in
Fig.~\ref{fig1}. At $c/a=1$ each Fe (Co) atom has 8 nearest neighbor
Co (Fe) atoms and 6 next nearest neighbor Fe (Co) atoms, etc. With
tetragonal distortion the distances between Fe or Co atoms in the
atomic plane become different compared to the adjacent planes in the
direction of the $c/a$ distortion and as a consequence the
intra-sublattice Fe-Fe and Co-Co exchange parameters split into two
components, which are denoted as $J_1$, $J_3$, $J_5$ and $J_2$, $J_4$
for in-plane and neighboring-plane parameters, respectively. As seen
from Fig.\,\ref{fig1} the nearest ($J_1$) and next-nearest neighbor
($J_2$) Fe-Fe (Co-Co) parameters stand out and are much affected by
the distortion showing in part variations between ferromagnetic and
antiferromagnetic values. A similar behavior is observed in the case
of Ni$_2$MnGa and hcp Gd under distortion.\cite{Ni2MnGa}

The most decisive role for $T_{\mathrm{C}}$ is played by the nearest
neighbor Fe-Co exchange as can be seen from Fig.\,\ref{fig1}(c):
firstly, their value is significantly larger than the value of the
Co-Co or Fe-Fe parameters shown in Fig.\,\ref{fig1}(a,b) [note the
different scale in Fig.\,\ref{fig1}(c) compared to
Fig.\,\ref{fig1}(a,b)]; secondly, we witness that they closely follow
the monotonical reduction of $T_{\mathrm{C}}$ with increasing
distortion, except for a flattening-out of $T_{\mathrm{C}}$ close to
$c/a=\sqrt{2}$ which is not followed by the Fe-Co interaction and
which we comment on later. The $c/a$ behavior of the exchange
interactions and resulting reduction of $T_{\mathrm{C}}$ can be
attributed to the complex exchange coupling mechanisms and will be
briefly discussed below.

Many-body model Hamiltonian approaches relevant to the problem provide
useful insight into the qualitative interpretation of the
density-functional results although a quantitative analysis of
exchange parameters $J$ in terms of different contributions is
frequently not possible. It is meaningful to separate the interaction
in two terms, $J=J_{\textrm{direct}} +
J_{\textrm{indirect}}$. $J_{\textrm{direct}}$ stems from the overlap
of $3d$ wavefunctions of neighboring atoms and practically vanishes
for distances larger than second-nearest neighbors. For FeCo it is
ferromagnetic because of the double-exchange mechanism\cite{double},
i.e., energy gain by broadening of the half-filled minority $d$ states
due to hybridization if the moments are parallel-aligned.
$J_{\textrm{indirect}}$ is mediated by the Fermi sea, concerns
interatomic distances from second-nearest neighbors and beyond, and is
analyzed here in terms of the Anderson \emph{s-d} mixing model because
of the localized nature of magnetic moments in these systems.

We proceed with a qualitative analysis of the calculated exchange
parameters. The monotonous reduction of the nearest neighbor Fe-Co
exchange interaction [see Fig.\,\ref{fig1}(c)] with tetragonal
distortion can be attributed to decrease of the direct coupling
$J_{\textrm{direct}}$ caused by three factors.  (i) The energetic
distance of the minority $d$ bands to the Fermi level decreases with
increasing $c/a$, from $~1$ eV to $~0.3$ eV, as seen in
Fig.~\ref{fig2}, leading to a strengthening of the antiferromagnetic
kinetic-exchange\cite{double} contribution to
$J_{\textrm{direct}}$. The latter mechanism is related to a repulsion
of the occupied majority-spin with unoccupied minority-spin levels of
neighboring atoms that stems from hybridization if the moments are
antiparallel-aligned and results in energy gain as the occupied levels
move lower in energy.  (ii) The inter-atomic Fe-Co distance increases
from 2.49\AA\ to 2.56\AA\ as $c/a$ increases from 1 to $\sqrtwo$
resulting in a weakening of the overlap of neighboring 3\emph{d} wave
functions. (iii) The magnetic moment amplitudes, that are included in
the values of $J_{ij}$ in Eq.~(\ref{eq:1}), decrease from 2.86
$\mu_\mathrm{B}$ to 2.65 $\mu_\mathrm{B}$ for Fe and from 1.82
$\mu_\mathrm{B}$ to 1.65 $\mu_\mathrm{B}$ for Co.

Concerning the indirect coupling, within the Anderson \emph{s-d}
mixing model $J_{\textrm{indirect}}$ can be separated into two
contributions (see, e.g., Ref.\,\onlinecite{RKKY}): $J_{\textrm{indirect}} =
J_{\textrm{RKKY}}+J_{\textrm{S}}$. Here the first term is an
oscillating Ruderman-Kittel-Kasuya-Yosida (RKKY)-like, which stems
from a spin polarization of the conduction electron sea by the local
moments. The second ``superexchange'' term, $J_{\textrm{S}}$, is
antiferromagnetic, decays exponentially with spatial distance, and
stems from virtual excitations in which electrons from local \emph{d}
states of Fe and Co are promoted above the Fermi sea. $J_{\textrm{S}}$
depends mostly on the distance of the unoccupied Fe (Co) 3\emph{d} peaks 
from the Fermi energy. The closer the peaks to the Fermi level, the stronger 
becomes $J_{\textrm{S}}$.

The intra-sublattice, i.e.\ Fe-Fe and Co-Co, exchange interactions
depend strongly on $J_{\textrm{indirect}}$, showing in part stronger
variations with tetragonal distortion compared to the Fe-Co
interaction. However, as the nearest neighbor Fe-Fe and Co-Co pairs
are relatively close (the in-plane distance decreases from 2.87\AA\ at
$c/a=1$ to 2.56\AA\ at $c/a=\sqrtwo$), the variation of $J_1$ also has
a direct-exchange contribution. In both sublattices we observe that
$J_1$ is sizeable and changes sign in the $c/a$ interval. In a large
section of the interval the various intra-sublattice interactions
partly compensate each other due to sizeable antiferromagnetic $J_1$
and $J_5$ (for Co-Co) terms. Only close to the end of the interval at
$c/a=\sqrt{2}$ do the Co-Co and Fe-Fe interactions contribute towards
a stronger ferromagnetic coupling, which results in a flattening-off
of the curve of $T_{\mathrm{C}}$ close to $c/a=\sqrt{2}$ that is not
witnessed in the Fe-Co coupling [see Fig.~\ref{fig1}(c)].

Having established the possibility of tuning $T_{\mathrm{C}}$ via the
$c/a$ ratio, we should note that an additonal parameter that may be
used for the tuning is the film thickness. It is known that two-dimensional
Heisenberg magnets without anisotropy have $T_{\mathrm{C}}=0$
\cite{Mermin}. However, in the presence of uniaxial MAE,
$T_{\mathrm{C}}>0$ and it grows with increasing film thickness, coming
close to the bulk value already at 15-20 atomic layers\cite{Erickson91,Tc_n1,Tc_n2} 
(depending of course on the magnitude of
$K_{\rm u}$). Since in HAMR applications one could conceivably wish a lower
$T_{\mathrm{C}}$ than the bulk limit shown here, this can be achieved
by reducing the film thickness. Of course the functionality will be
also determined by the thickness dependence of $K_{\rm u}$ which we do not
study here, however, for thin films $K_{\rm u}$ is expected to be
appreciable because of the reduced symmetry even if it differs from
the value of $600~\mu$eV that was found for $c/a=1.24$.

In conclusion, combining density-functional theory calculations with a
classical Monte Carlo method, we show that for B2-type FeCo compounds
a tetragonal distortion with $1<c/a<\sqrt{2}$ gives rise to a strong
reduction of the Curie temperature $T_{\mathrm{C}}$. In this interval
the $T_{\mathrm{C}}$ decreases monotonically from 1575 K to 940 K. We
find that due to the strong ferromagnetic character of FeCo compounds
the exchange interactions are strongly damped for large interatomic
distance and thus the nearest neighbor Fe-Co exchange interaction is
sufficient to explain the $c/a$ dependence of the
$T_{\mathrm{C}}$. Combination of high magnetocrystalline anisotropy
energy with a moderate $T_{\mathrm{C}}$ value suggests tetragonal FeCo
grown on the Rh substrate with $c/a=1.24$ to be a promising material
for HAMR applications.

This work was partly supported by the Young Investigators Group
Programme of the Helmholtz Association, Germany, contract
VH-NG-409. We gratefully acknowledge the support of J\"ulich
Supercomputing Centre.

\end{document}